\documentclass{article}
\usepackage{spconf}
\usepackage{amsmath}
\usepackage{amssymb}
\usepackage{graphicx}
\usepackage[hidelinks]{hyperref}
\usepackage{siunitx}
\usepackage[vietnamese]{babel}
\usepackage[caption=false,font=footnotesize]{subfig}
\usepackage[export]{adjustbox}

\DeclareMathOperator*{\argmin}{arg\,min}

\def \akis [#1]{\textcolor{red}{AP: #1}}  

\title{Position tracking of a varying number of sound sources with \\ sliding permutation invariant training}
\name{David Diaz-Guerra$^{\star}$ \sthanks{This work was realized during a research stay at the Audio Research Group of the Tampere University (Finland) supported in part by the University of Zaragoza and \textit{Fundación Bancaria Ibercaja y Fundación CAI} (ref. number IT 8/22). This work was supported in part by the Regional Government of Aragon (Spain) with a grant for postgraduate research contracts (2017-2021) co-founded by the Operative Program FSE Aragon 2014-2020.}
\qquad Archontis Politis$^{\dagger}$ 
\qquad Tuomas Virtanen$^{\dagger}$}

\address{$^{\star}$ Dept. of Electronic Engineering and Communications, University of Zaragoza, Spain \\
  $^{\dagger}$Audio Research Group, Tampere University, Finland}

\begin{document}
%\ninept
%
\maketitle
\begin{abstract}
%A multi-source tracking system must not only provide position or direction of arrival (DOA) estimates but also detect when a new source appears or disappears and being able to associate every new estimation to the correct tracked trajectory avoiding identity switches (IDSs). In addition, when we cannot apply any criteria to order or classify the sound sources, it becomes a problem invariant to the permutation of the sources. In this paper, we present a new training strategy for deep learning models that is able to optimize all the desired properties of a tracking system, having a simple expression based on the Mean Squared Error and being easy to implement.
Machine-learning-based sound source localization (SSL) methods have shown strong performance in challenging acoustic scenarios. However, little work has been done on adapting such methods to track consistently multiple sources appearing and disappearing, as would occur in reality. In this paper, we present a new training strategy for deep learning SSL models with a straightforward implementation based on the mean squared error of the optimal association between estimated and reference positions in the preceding time frames. It optimizes the desired properties of a tracking system: handling a time-varying number of sources and ordering localization estimates according to their trajectories, minimizing identity switches (IDSs). Evaluation on simulated data of multiple reverberant moving sources and on two model architectures proves its effectiveness in reducing identity switches without compromising frame-wise localization accuracy.
\end{abstract}
\section{Introduction}
\label{sec:intro}

Machine-Learning-based sound source localization (SSL) using neural networks (NNs) has been an intensely active research field during the last years \cite{grumiaux2022} but with a fair amount of problems still open. A major one is how to deal with the common real-world scenario that the number of active sources in the scene varies, as sources appear and disappear dynamically, with proposed solutions employing either simultaneous classification \cite{adavanne2019} or assigning temporal source activity probabilities for a fixed number of model outputs \cite{schymura2021exploiting_etal, adavanne2021}. Another closely related open problem is associating new location estimates to the appropriate model output in the case of multiple simultaneous sources, such that the same output provides a consistent sequence of locations tracking the trajectory of the same source. The above combined tracking objectives are essential in making deep-learning localizers useful in further downstream tasks, such as beamforming, source separation, or robot audition, that can utilize the location information of multiple sources in the scene.
%Sound source localization (SSL) using neural networks has been an intensely active research field during the last years \cite{grumiaux2022} but with a fair amount of questions still open. Among them, it is the question about how to deal with the invariance to the permutation of the sources when we cannot use any criteria to order or classify them and therefore we cannot expect the model to output the tracked trajectories in the same order as they are in the ground-truth data of our training dataset. In addition, when we want to move from SSL to sound source tracking (SST) we need to, apart of estimate the position of every source, be able to properly detect when they appear or disappear and assign every new detection to the correct trajectory without switching their identities. 

%Most of the deep-learning based SSL and SST systems proposed in the literature work on a sound event localization and detection (SELD) framework where only one source of every class can be simultaneously active. Under this framework, detecting the sound events and keeping the identities of every sound source stable becomes just a multi-label classification problem which, in addition, is usually solved in a separated branch; i.e., the models have a branch with a multi-label classification output to detect the sound events and another branch which continuously provides localization estimates for every event class and that is only taken into account when the detection branch detect an event of that class.

Multi-source localization has been tackled in the literature mainly using a classification framework, e.g. \cite{chakrabarty2019, perotin2019a}, indicating source presence probabilities on a grid of locations, which can naturally accommodate a time-varying number of sources but without providing assignment between detected spatial labels and source identities or trajectories.  Proposals that localize multiple sources using regression follow mainly a simultaneous localization and detection framework (SELD) \cite{politis2020overview, shimada2021} which naturally associates a source trajectory to temporal activity of a target class, but only in the case that there are no multiple sources of the same class active simultaneously. A few regression-based localization works have aimed to tackle that problem using a frame-wise permutation invariant training (PIT) strategy, either in a SELD context \cite{cao2021improved_etal, shimada2022multi_etal} or in a pure SSL one \cite{krause2021data, subramanian2022deep_etal}. PIT allows effective training of regression-based multi-source localization, but without promoting the tracking objectives outlined above.

%In \cite{shimada2021}, the use of activity-coupled Cartesian DOA (ACCDOA) vectors ---vectors steering at the estimated DOA and whose length represents their probability of corresponding to an active source--- is proposed as a way to encode both the localization and the detection information at the output of the models. This representation is quite convenient since minimizing their mean squared error (MSE) simultaneously optimize both the localization and detection capabilities of the models. However, they used these ACCDOA vectors in a sound event localization and detection (SELD) context with the restriction of having a maximum of one source of each class simultaneously active, so they could use an ACCDOA output for every class without needing to worry about the invariance of the sources or not switching the identities of the sources.

%To solve the invariance problem, the permutation invariant training (PIT) strategy is used in \cite{subramanian2022deep} to train models able to localize multiple simultaneous talkers. The PIT chooses, for every time frame, the permutation of the estimated sources that minimizes the assignment cost between them and the ground-truth trajectories and then backpropagates the Mean Squared Error resulting from these assignments. This approach can be used to train SSL models, but it does not penalize the identity switches (IDSs) so they need to be fixed in post-processing stages if we want to have a full SST system.

A recent attempt aiming to optimize directly tracking objectives for SSL models is presented in \cite{adavanne2021}, where a NN is trained to infer the frame-wise assignments between estimated and ground truth directions-of-arrival (DOAs), which are further used to construct differentiable versions of multi-object-tracking metrics \cite{bernardin2008}. Based on those, tracking losses are back-propagated to the outputs of the localization model during its training. While the method is shown to be effective, using an auxiliary NN to compute the loss function adds implementation and training complexity, with harder-to-interpret gradients that are more prone to vanishing effects.
%apart of being harder to implement and having to re-train this network in every new setup if we want to train new models in different scenarios.

%A different approach is followed in \cite{adavanne2021} where, borrowing the idea from the computer vision field \cite{xu2020}, they propose a pre-trained neural network to solve the assignment problem between the estimated and the ground-truth DOAs that they use to define differentiable versions of the CLEAR MOT metrics \cite{bernardin2008}. They used this approach to train a SST model that was able to track up to 2 simultaneous sources but, using an auxiliary neural network to compute the loss function, made its gradients harder to interpret and increase the risk of them to vanish, apart of being harder to implement and having to re-train this network in every new setup if we want to train new models in different scenarios.

 In this paper, we propose a modification to PIT that penalizes IDSs at each frame by considering the source permutation that minimizes the MSE over a number of preceding frames.
 %, by using in every time frame the source permutation that minimizes the MSE in the previous frames instead of only in the current one, penalizes IDSs made by the model. In addition, if used over ACCDOA vectors, it also minimizes the detection misses and false positives, so it optimizes all the requirements of a SST system.
 Additionally, we show that if the SSL model provides DOAs in the \emph{activity-coupled} DOA (ACCDOA) representation \cite{shimada2021}, which joins localization and detection information by scaling a DOA vector by its probability of belonging to an active source, then the proposed training also minimizes source misses and false positives. Hence, it makes the model reactive to variable source conditions and optimizes all the tracking objectives outlined earlier.
 This new PIT strategy is easy to implement, has a very similar computational cost to the standard PIT and, being based on backpropagating the MSE of one of the possible permutations of the estimated DOAs, it generates gradients that are strong and easy to interpret.

\section{Permutation Invariant Training for sound source tracking}
\label{sec:PIT}

When we want to train a sound source tracking (SST) model in a supervised manner and we cannot apply any criteria to classify and order the sources, we have to face the permutation invariance of the sources; i.e., we cannot directly compare the m-th trajectory estimated by the model $\hat{\mathbf{y}}_m(t)$ with the m-th trajectory of our ground-truth dataset $\mathbf{y}_m(t)$ since the model cannot infer the ground-truth order of the trajectories. This is a well-known issue in the speech separation field, where permutation invariant training (PIT) was first proposed \cite{yu2017a}.

All the PIT strategies propose finding a permutation $\sigma: m \rightarrow \sigma_m, \; \forall m \in \{0, ..., M-1\}$ according to certain optimization criteria to reorder the outputs of the neural network and then use it to compare the estimated and ground-truth trajectories. When using ACCDOA vectors to represent the DOA and the activity of the sources, we can use the mean squared error (MSE) as the loss function to train our models:
\begin{equation}
    \label{eq:PITLoss}
    L_{PIT} = \frac{1}{TM} \sum_{t=0}^{T-1} \sum_{m=0}^{M-1} \lVert \mathbf{y}_m(t) - \hat{\mathbf{y}}_{{\sigma}_m(t)}(t) \rVert^2
\end{equation}
where $M$ is the maximum number of trajectories that the model can estimate, $T$ is the number of time frames in the scene, and $\rVert \cdot \rVert$ is the Euclidean norm operator. In the case of having a number of ground-truth trajectories lower than $M$, we can just add as many 0-norm padding trajectories.

\begin{figure}[tb]
    \centering
    \includegraphics[width=0.66\linewidth]{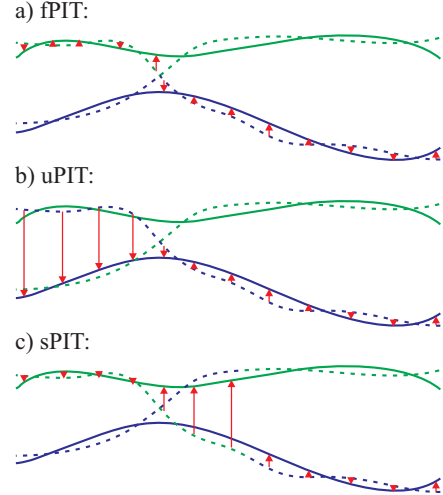}
    \caption{Examples of 1D trajectories and the result of applying the different PIT strategies. The dashed lines represent the estimates and their color how every PIT pair them with the ground-truth trajectories (solid lines). The red arrows represent the gradients of the MSE of every pairing w.r.t. the first estimated trajectory.}
    \label{fig:PITs}
\end{figure}

\subsection{Frame-level Permutation Invariant Training (\MakeLowercase{f}PIT)}

The original PIT \cite{yu2017a} applied to SSL \cite{krause2021data, subramanian2022deep_etal}, which we will call frame-level PIT (fPIT), finds the permutation of the estimated sources that minimizes the matching error between the estimated and the ground-truth DOAs for every time frame:
\begin{equation}
\label{eq:fPIT}
    \sigma^f(t) = \argmin_{\sigma \in \Pi_M} \sum_{m=0}^{M-1} \lVert \mathbf{y}_m(t) - \hat{\mathbf{y}}_{\sigma_m}(t) \rVert
\end{equation}
where $\Pi_M$ is the set of all the permutations $\sigma: i \rightarrow \sigma_i$ of $M$ elements. This optimization problem can be solved by computing the $M \times M$ distance matrices $\mathbf{D}(t)$ with elements $d_{ij}(t) = \lVert \mathbf{y}_i(t) - \hat{\mathbf{y}}_j(t) \rVert$ and applying the Hungarian algorithm over them to find the optimal permutation for every time frame.

Using $\sigma^f(t)$ in \eqref{eq:PITLoss} allows us to solve the permutation invariance problem but, as we can see in Fig. \ref{fig:PITs}a, fPIT does not penalize at all the IDSs. Instead, its gradients push the model to do these switches as fast as possible so that estimations are close to a ground-truth trajectory at all time frames. Therefore, if we want to keep the identity of every output stable during tracking, we need to add post-processing stages to fix the IDSs.

\subsection{Utterance-level Permutation Invariant Training (\MakeLowercase{u}PIT)}

In order to penalize the IDSs, utterance-level PIT (uPIT) \cite{kolbaek2017} proposes finding the permutation that minimizes the error for a whole speech utterance or some other longer recording unit of interest, instead of a different one for every time frame:
\begin{equation}
\label{eq:uPIT}
    \sigma^u = \argmin_{\sigma \in \Pi_M} \sum_{t=0}^{T-1} \sum_{m=0}^{M-1} \lVert \mathbf{y}_m(t) - \hat{\mathbf{y}}_{\sigma_m}(t) \rVert
\end{equation}
where $T$ is the number of time frames of the acoustic scene. In this case, we only need to apply the Hungarian algorithm once per acoustic scene after computing the time average of the matrices $\mathbf{D}(t)$.

Replacing $\sigma^f(t)$ by $\sigma^u$ in \eqref{eq:PITLoss} indeed penalizes the presence of IDSs since all the frames where the output ACCDOAs do not follow the main identity assignment count as completely wrong estimations. However, as we can see in Fig. \ref{fig:PITs}b, this penalization is excessive, being able to penalize situations that can not be solved by causal systems or generating gradients too much time after the IDS when, in most situations, it would be preferred to focus and keep tracking on the new identities rather than switching them again. Our experiments in using uPIT for multi-source tracking showed that it can easily generate a wide local minimum in the loss function that corresponds to estimating all DOAs in the middle of the active sources. That effect makes effective model training impossible, especially for long scenes of variable multiple sources.

\section{Sliding Permutation Invariant Training (\MakeLowercase{s}PIT)}
\label{sec:sPIT}

To overcome the limitations of both fPIT and uPIT, we propose a new PIT strategy that we call sliding PIT (sPIT) which consists in choosing, for every time frame, the optimal permutation for the last $T_{avg}$ frames, i.e., for a causal sliding window of length $T_{avg}$:
\begin{equation}
\label{eq:sPIT}
    \sigma^s(t) = \argmin_{\sigma \in \Pi_M} \sum_{k=0}^{T_{avg}-1} \sum_{m=0}^{M-1} \lVert \mathbf{y}_m(t-k) - \hat{\mathbf{y}}_{\sigma_m}(t-k) \rVert
\end{equation}

In order to obtain $\sigma^s(t)$ for every time frame, we can follow the same procedure as in the fPIT but applying a causal moving average of length $T_{avg}$ over the elements of $\mathbf{D}(t)$ before computing the Hungarian algorithm, so the computational complexity is virtually the same. It is worth mentioning that, in the case of training non-causal trackers, we could replace the causal moving window in \eqref{eq:sPIT} by a centered window.

%In order to obtain $\sigma^s(t)$ for every time frame, we compute a $M \times M$ distance matrix $\mathbf{D}(t)$ with elements $d_{ij}(t) = \lVert \mathbf{y}_i(t) - \hat{\mathbf{y}}_j(t) \rVert$, apply a causal moving average of length $T_{avg}$ over their elements, and finally use the Hungarian Algorithm over the resulting matrix at every time frame.

As shown in Fig. \ref{fig:PITs}c, the sPIT penalizes an estimation if it does not follow the main source assignment of the last $T_{avg}$, while it stops penalizing an IDS after a maximum of $T_{avg}$ frames and
%, from that point, it 
focuses on maintaining the new identities. Hence, the global minimum of the loss function corresponds to a solution without any IDSs that also avoids the training converging to the useless local minima generated by uPIT, estimating all DOAs in the middle point of the active sources.

In addition, when used over ACCDOA vectors, if the number of estimated sources is lower than the actual number, one of the estimated ACCDOA vectors whose norm is lower than the detection threshold will be paired with the ground-truth ACCDOA vector of the missed source and the gradients of \eqref{eq:PITLoss} will pull that estimated ACCDOA towards it. In a similar manner, in the case of a false positive, the gradients will pull the false-positive ACCDOA towards 0. Hence, sPIT is able to optimize both source detections and consistent source assignments that we expect from a competent SST system.

\section{Evaluation}
\label{ref:evaluation}

\subsection{Experiment design}

To evaluate sPIT, we have trained and evaluated a fully convolutional model and a convolutional model with recurrent layers at its end using simulated scenarios with up to three active sources at the same time. We trained the models using fPIT, uPIT, and sPIT, but we do not include the uPIT results since it did not converge to any proper solution.

We have used a synthetic dataset similar to the one used in \cite{diaz-guerra2022c}, but with the number of sources varying during each \SI{20}{\s} scene. Every \SI{200}{\ms} new sources could birth with a rate of 0.06, 0.04, or 0.02 depending on whether there are already 0, 1, or 2 active sources. Once they are born, they could have a minimum duration of \SI{2}{\s} and, after that, every \SI{200}{\ms} they could die with a probability of 0.02.

For each trajectory, we randomly chose a starting and ending point and connect them using sinusoidal functions in the three Cartesian coordinates and simulate the room acoustics in a reverberation range from $T_{60}=$ 0.2 to \SI{1.3}{\s} using the Image Source Method \cite{allen1979} with utterances from the LibriSpeech dataset \cite{panayotov2015} as source signals and a 12-microphone array designed to be mounted over a NAO robot head \cite{lollmann2018_etal} as receiver. % $T_{60}=\SIrange{0.2}{1.3}{\s}$

%We have used two models, one fully convolutional and one including recurrent layers at its end. 

\begin{figure}[tb]
    \centering
    \includegraphics[width=0.75\linewidth]{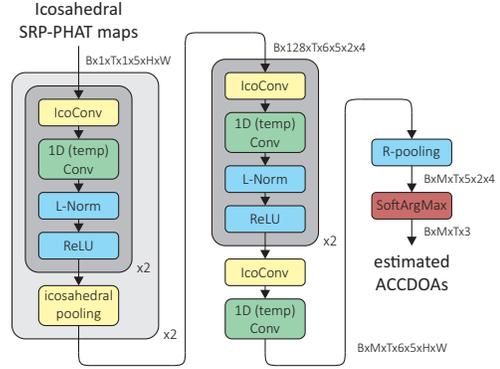}
    \caption{Architecture of the icoCNN used for evaluation. B is the batch size, T is the number of temporal frames of the acoustic scenes, $H=2^r=8$ and $W=2^{r+1}=16$ are the height and the width of the projections of the icosahedral grid.}
    \label{fig:icoCNN}
\end{figure}

\begin{figure}[tb]
    \centering
    \includegraphics[width=0.75\linewidth]{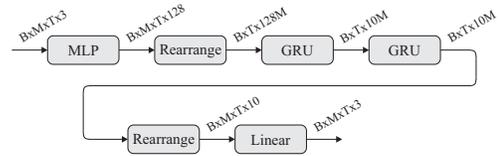}
    \caption{Architecture of the RNN used after the icoCNN for evaluation.}
    \label{fig:RNN}
\end{figure}

\begin{figure*}[tb]
    \centering
    \subfloat{%[\label{fig:MAE}]{
        \includegraphics[width=0.2\linewidth,valign=c]{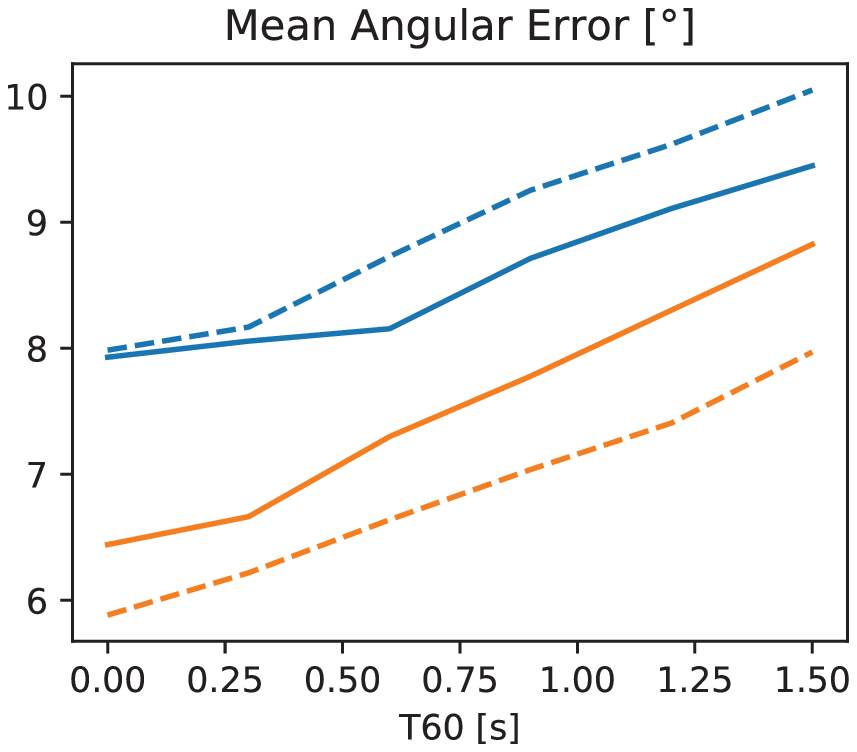}
    }
    \subfloat{%[\label{fig:IDSs}]{
        \includegraphics[width=0.2\linewidth,valign=c]{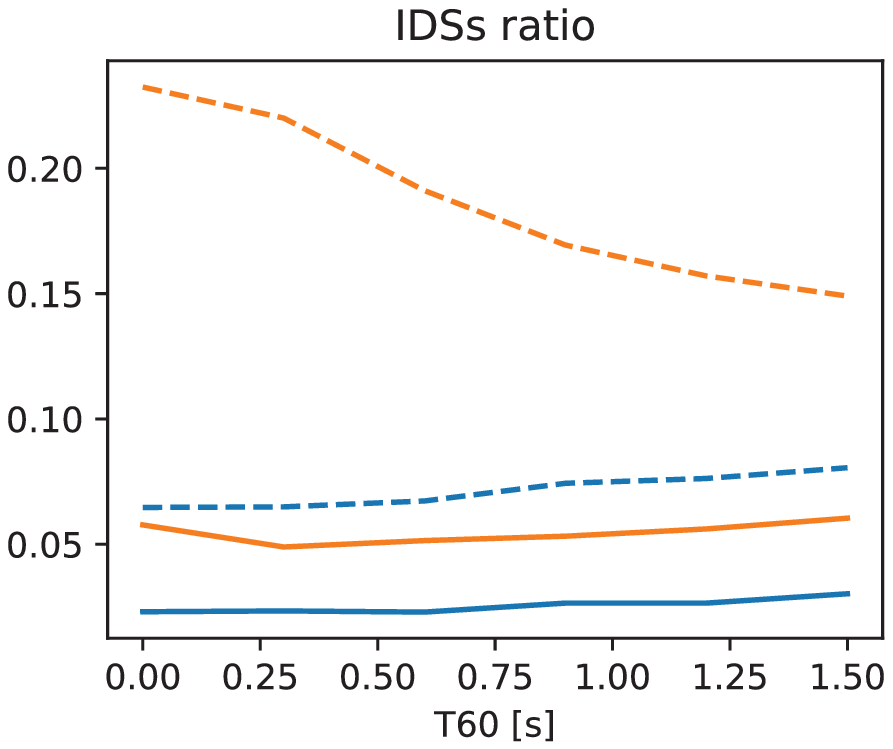}
    }
    \subfloat{%[\label{fig:ROC}]{
        \includegraphics[width=0.22\linewidth,valign=c]{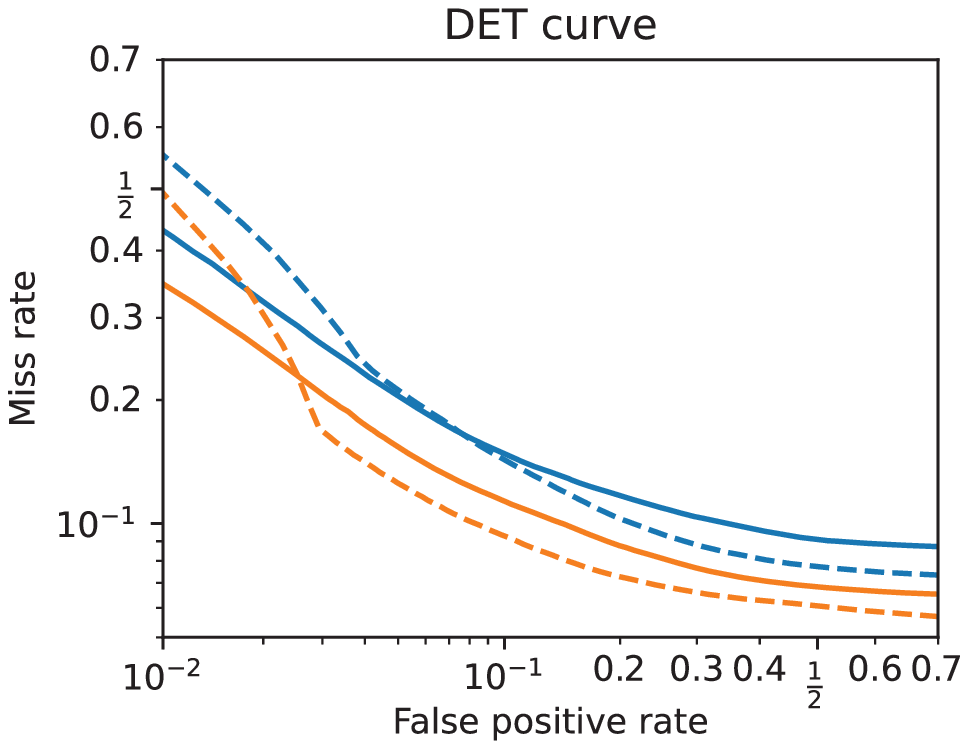}
    }
    \subfloat{
        \includegraphics[width=0.16\linewidth,valign=c]{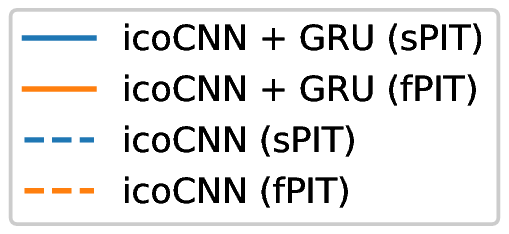}
    }\hfill
    \caption{Evaluation metrics obtained using fPIT and sPIT to train the two analyzed models.}
    \label{fig:results}
\end{figure*}

% \begin{figure*}[tb]
%     \centering
%     \subfloat{%[\label{fig:example_sPIT}]{
%         \includegraphics[width=0.33\linewidth]{images/example_01_sPIT_small.eps}
%         %\caption{Example of an evaluation acoustic scene and output of the icoCNN + GRU model trained with sPIT.}
%     }
%     \subfloat{%[\label{fig:example_fPIT}]{
%         \includegraphics[width=0.33\linewidth]{images/example_01_fPIT_small.eps}
%         %\caption{Example of an evaluation acoustic scene and output of the icoCNN + GRU model trained with fPIT.}
%     }
%     \caption{Example of an evaluation acoustic scene and output of the icoCNN + GRU model trained with sPIT and fPIT.}
%     \label{fig:examples}
% \end{figure*}

\begin{figure}[tb]
    \centering
    \subfloat{%[\label{fig:example_sPIT}]{
        \includegraphics[width=0.5\linewidth]{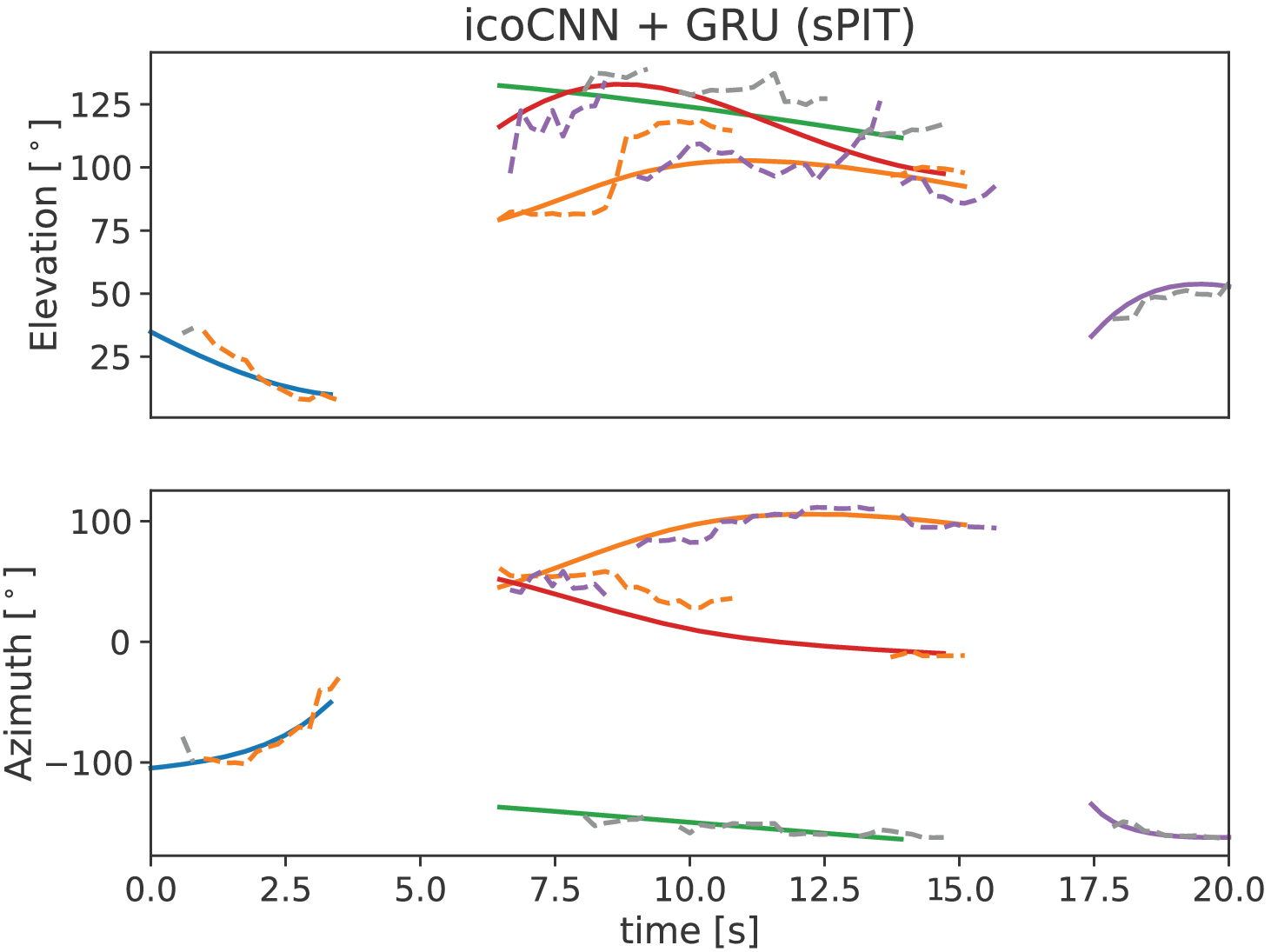}
        %\caption{Example of an evaluation acoustic scene and output of the icoCNN + GRU model trained with sPIT.}
    }
    \subfloat{%[\label{fig:example_fPIT}]{
        \includegraphics[width=0.5\linewidth]{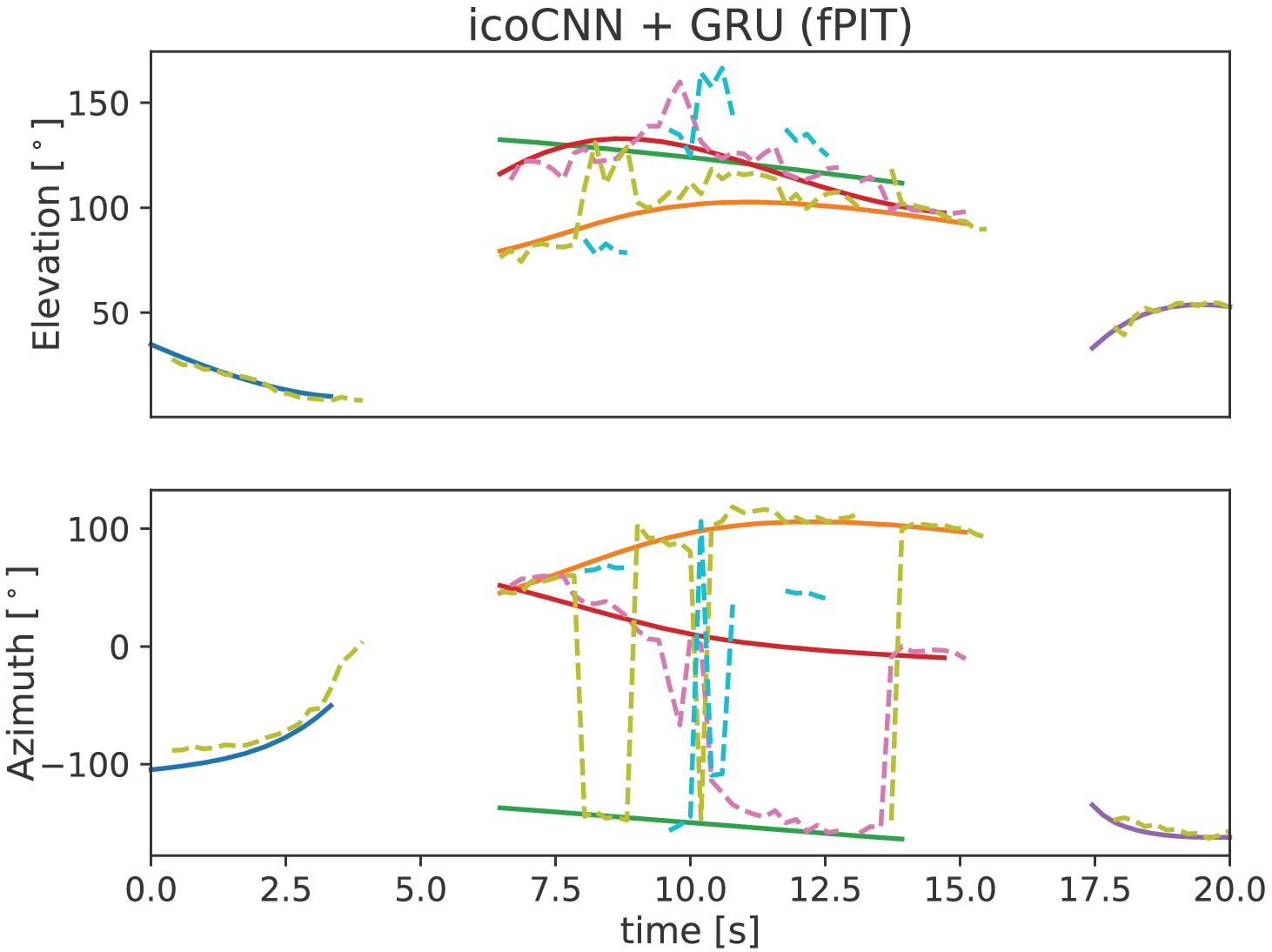}
        %\caption{Example of an evaluation acoustic scene and output of the icoCNN + GRU model trained with fPIT.}
    }
    \caption{Example of an evaluation acoustic scene and output of the icoCNN + GRU model trained with sPIT and fPIT. The solid lines represent the ground-truth trajectories and the dashed lines the estimations.}
    \label{fig:examples}
\end{figure}

% \begin{figure}[tb]
%     \centering
%     \subfloat{%[\label{fig:example_sPIT}]{
%         \includegraphics[width=0.70\linewidth]{images/example_01_sPIT_small.eps}
%         %\caption{Example of an evaluation acoustic scene and output of the icoCNN + GRU model trained with sPIT.}
%     } \\
%     \subfloat{%[\label{fig:example_fPIT}]{
%         \includegraphics[width=0.70\linewidth]{images/example_01_fPIT_small.eps}
%         %\caption{Example of an evaluation acoustic scene and output of the icoCNN + GRU model trained with fPIT.}
%     }
%     \caption{Example of an evaluation acoustic scene and output of the icoCNN + GRU model trained with sPIT and fPIT.}
%     \label{fig:examples}
% \end{figure}

For the fully convolutional model, we have modified the model presented in \cite{diaz-guerra2022c} to make it work in multi-source scenarios. The original model computed SRP-PHAT power maps defined on a tessellated icosahedron using frames of 4096 samples with a sampling rate of \SI{16}{\kHz} and analyzed them with an icosahedral CNN (icoCNN) \cite{cohen2019} using 1D temporal convolutions for tracking. The icoCNN transformed the power maps into an icosahedral probability distribution whose expected value, computed with a Soft-ArgMax layer, corresponded to the estimated DOA. In order to adapt the model to handle the multi-source case, we just increased the number of channels of the last convolutional layer from 1 to $M$. The general architecture of this model is represented in Fig. \ref{fig:icoCNN}, with all the convolutions having 128 channels except for the last temporal convolution having $M$. We used input maps of resolution $r=3$ (i.e., with 630 grid points) and temporal convolutions with kernels of size 5, so the model had a temporal receptive field of \SI{7.17}{\s}. %Since the Soft-ArgMax layer already operated with 3D unit vectors pointing at every grid point, its outputs can already be interpreted as ACCDOA vectors whose norm is related to the variance of the probability distribution generated by the icoCNN. Therefore, in order to use this model in a multi-source scenario, we just needed to use $M$ output channels instead of only one to obtain $M$ ACCDOA vectors.

Since most of the SST models in the literature include recurrent layers after their CNNs \cite{adavanne2019, perotin2019a, adavanne2021, shimada2021, krause2021data}, we also evaluate the same model adding the RNN depicted in Fig. \ref{fig:RNN} at its end. It uses a multi-layer perceptron (MLP) to expand the 3D ACCDOA vectors generated by the convolutional model to a space of 128 dimensions. It then concatenates the $M$ representations of every trajectory and applies two gated recurrent units (GRUs) 
%\cite{cho2014a} 
to finally split the output of the last GRU again into $M$ elements representing every tracked trajectory. Finally, a linear projection layer is used to obtain the 3D ACCDOA vectors of every source. With this model, apart from the evaluated PIT over the final ACCDOAs, we also included a fPIT loss over the ACCDOAs generated by the convolutional part in order to facilitate its training.

We trained models using the AdamW algorithm \cite{loshchilov2018} with gradient clipping during 100 epochs. We used $M=10$ as the number of ACCDOA outputs of all our models since we observed that it was beneficial to use a higher number than the maximum possible number of active sources in the dataset (i.e., 3). All the results shown in this paper were obtained using $T_{avg}=10$ frames (i.e., \SI{2}{\s}). No large changes were observed for windows in the range $T_{avg}\in[5,20]$. We also tried to train the models using the dMOT approach proposed in \cite{adavanne2021}, but we could not train their Hungarian network to solve the assignment problem with $M=10$. Still, it is worth saying that this might be possible with other network architectures \cite{kameoka_attentionpit_2022}.

\subsection{Results}

% \begin{figure}[tb]
%     \centering
%     \includegraphics[width=0.7\linewidth]{images/example_01_sPIT_small.eps}
%     \caption{Example of an evaluation acoustic scene and output of the icoCNN + GRU model trained with sPIT.}
%     \label{fig:example_sPIT}
% \end{figure}

% \begin{figure}[tb]
%     \centering
%     \includegraphics[width=0.7\linewidth]{images/example_01_fPIT_small.eps}
%     \caption{Example of an evaluation acoustic scene and output of the icoCNN + GRU model trained with fPIT.}
%     \label{fig:example_fPIT}
% \end{figure}

Fig. \ref{fig:results} shows the evaluation results in terms of the mean angular error (MAE) of the true positives (TP), the ratio between the number of IDSs and the number of ground-truth objects, and the relation between the miss and false positive (FP) ratios when tuning the value of the detection threshold. The MAE and the IDSs ratio were computed using 0.5 as the detection threshold on the norm of the output ACCDOA vectors and any estimate with a localization error higher than \ang{30} was considered a FP and a miss rather than a TP with high error.

For the localization accuracy, we can see how both models have a similar MAE with both training strategies. For the fully convolutional model, the sPIT degrades the MAE by about \ang{2} and in the model using GRUs only by \ang{1}. Similarly, we do not observe big differences in the DET curves, with the models trained with sPIT having slightly higher miss ratios for the same level of FPs. However, we can observe big differences in terms of IDSs, with sPIT reducing the IDSs ratio by more than a factor of 2 in the model using recurrent layers. Even with the fully convolutional network that does tracking with only convolution operations over the previous \SI{7.17}{\s} of the trajectory, sPIT is able to keep the IDSs ratio under 0.1. Finally, Fig. \ref{fig:examples} shows an example of an acoustic scene and the tracking obtained using sPIT and fPIT. Even if it still has some IDSs when 3 sources are simultaneously active, the model trained with sPIT has much fewer IDSs than the one trained with fPIT.

\section{Conclusions}
\label{sec:conclusions}

We have presented a sliding PIT strategy that is able to strongly reduce the number of IDS of the SST models sacrificing just a small deterioration of the precision in the localization and of the compromise between precision and recall. The proposed sliding PIT will allow the training of SST models whose output can be used as tracking estimates without needing to add any post-processing steps such as assignment algorithms to avoid IDSs or peak-picking algorithms over classification outputs to choose the correct DOAs.

%The PI-GRUs did not show too important benefits in the presented experiments, but considering that these are the first attempts of using PI-RNNs, we believe that the results are promising. One of their main benefits, being able to store more information per source, is not fully exploited with this architecture (where only geometrical information is provided to them) and we could expect more benefits compared to the conventional GRUs by including spectral information of every tracked source. Further studies on the way that their state set is initialized would also be interesting.

\bibliographystyle{IEEEtran}
\bibliography{MicrophoneArraysRedux.bib}

% Below is an example of how to insert images. Delete the ``\vspace'' line,
% uncomment the preceding line ``\centerline...'' and replace ``imageX.ps''
% with a suitable PostScript file name.
% -------------------------------------------------------------------------
% \begin{figure}[htb]

% \begin{minipage}[b]{1.0\linewidth}
%   \centering
%   \centerline{\includegraphics[width=8.5cm]{image1}}
% %  \vspace{2.0cm}
%   \centerline{(a) Result 1}\medskip
% \end{minipage}
% %
% \begin{minipage}[b]{.48\linewidth}
%   \centering
%   \centerline{\includegraphics[width=4.0cm]{image3}}
% %  \vspace{1.5cm}
%   \centerline{(b) Results 3}\medskip
% \end{minipage}
% \hfill
% \begin{minipage}[b]{0.48\linewidth}
%   \centering
%   \centerline{\includegraphics[width=4.0cm]{image4}}
% %  \vspace{1.5cm}
%   \centerline{(c) Result 4}\medskip
% \end{minipage}
% %
% \caption{Example of placing a figure with experimental results.}
% \label{fig:res}
% %
% \end{figure}

%\vfill
%\pagebreak

\end{document}